# A bias-free quantum random number generation using photon arrival time selectively

**Jian-min Wang**[1,†], **Tian-yu Xie**[1,†], **Hong-fei Zhang**[1], **Dong-xu Yang**[1], **Chao Xie**[3] **and Jian Wang**[1,2,*], Senior Member, IEEE

1 State Key Laboratory of Technologies of Particle Detection and Electronics, University of Science and Technology of China, Hefei, Anhui, 230026, China
2 Synergetic Innovation Center of Quantum Information and Quantum Physics, University of Science and Technology of China, Hefei, Anhui 230026, China
3 Electrical & Information Engineering Department, Huang Shan University, Huangshan, Anhui 245021, China
† these authors equally contribute to this work
*wangjian@ustc.edu.cn

**Abstract:** We present a high-quality, bias-free quantum random number generator (QRNG) using photon arrival time selectively in accordance with the number of photon detection events within a sampling time interval in attenuated light. It is well showed in both theoretical analysis and experiments verification that this random number production method eliminates both bias and correlation perfectly without more post processing and the random number can clearly pass the standard randomness tests. We fulfill theoretical analysis and experimental verification of the method whose rate can reach up to 45Mbps.

## 1. Introduction

Random number generators are applied to a wide variety of traditional fields [1] like theoretical simulation in physics, Monte Carlo method, random walk, etc. Pseudo-random numbers obtained from a deterministic algorithm can meet the requirements in these fields where the existence of long period and determinism have no effects. Recently, along with the emergence of new fields [2] as quantum cryptography, quantum ID identification and so forth, demands for physical true random numbers is becoming more and more urgent. Because quantum processes are unpredictable in theory, random numbers derived from quantum processes could be called physical true random numbers.

It is not a short history to yield random numbers uing quantum process. Earlier in the 1970s, Frigerio [3] utilized nuclear decay to extract randomness, although the rate was low, they blazed a trail indeed. With the appearance of the scheme using beamsplitter [4-6] to bifurcate the path of a photon, people began to concentrate more on randomness extraction from optical field, mainly due to two advantages: it is not difficult to control and detect optical field; photons are so highly dense compared to alpha or beta particles in nuclear decay that we can obtain more entropy and achieve the rate order of Mbps. Photon arrival time[7-10], photon counts [11] within sampling intervals and random phase [12-13] essentially originate from random fluctuation of optical field. Researches in these three aspects have already been accumulated gradually in the last few years. Those schemes almost use post-processing methods to remove bias or whiten the raw data and pass the random number statistical tests. A better scheme is that post-processing should not be used especially some post-processing methods can whiten the raw data to pass the randomness test. There are some reports for post-processing free or bias-free QRNG such as [14] with a rate of 4.01 Mbps, [9] with a rate of 1Mbps and [15] with a rate of 8Mbps.

In this paper we demonstrate a bias-free random number production scheme with the high rate up to 45Mbps. We pay more attention to theoretical analysis and experimental verification and set up a high-quality bias-free quantum random number generator using photon arrival time selectively without more post-processing. In the following we illuminate our scheme in paragraphs, first explaining the theory of our design, then detailing its implementation as well as randomness tests performed, and finally discussing further improvements and outlook.

## 2. SETUP AND THEORY ANALYSIS

The set up of our experiment is shown in Fig.1, the light emitted by a light emitting diode (LED) is attenuated by the cross polarizers (CP) to the single photon level, and the illumination of LED is adjustable by an adjustable resistor. To achieve high rates of random numbers a photomultiplier tube (PMT) with high sensitivity is used for detecting the attenuated light. The pulses generated from PMT are amplified by AMP module and discriminated by a discriminator (DS) module which converts the analog output pulse of the PMT into a digital signal. The discriminator can distinguish two pulses only if two pulses are separated by about the pulse width, otherwise they will overlap and form only one pulse discriminated by the discriminator. This leads to an effective dead time $t_d$ which equals the width of pulses and is the total effect of PMT, AMP and discriminator. Output of the discriminator is fed into the time-digital convertor (TDC) module and FPGA(Field Programmable Gate Array) module, the TDC measures arrival time of all the pulses, relative to external reference clock with a frequency of $f_0$ Hz, followed by the FPGA that selects the period of only one pulse and outputs the arrival time of the selected pulse.

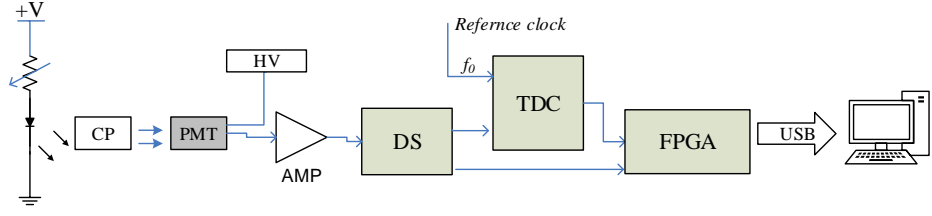

Fig.1 Schematic of our experimental setup. CP: Cross Polarizer, PMT: Photomultiplier, HV: High Voltage, AMP: Amplifier, DS: Discriminator, TDC: Time-Digital Convertor, FPGA: Field Programmable Gate Array, USB: Universal Serial Bus. Parameters of PMT: HV=1000V, Gain=1.1M, dark counts ~ 200, PMT type = R7400U

The stimulated emission of photons from a semiconductor device is believed to be a process in which events occur continuously and independent of each other, i.e., a Poissionan process. The probability to register $k$ clicks within a period of time T is given by Poissonian distribution:

$$P(X=k) = e^{-\lambda T}\frac{(\lambda T)^k}{k!} \qquad (1)$$

Where $\lambda$ is the average number of events per unit time. The distribution of arrival time of the first detection is calculated by

$$f(T=t) = \frac{\lambda e^{-\lambda t}}{\int_0^{T_0} \lambda e^{-\lambda t} dt} = \frac{\lambda e^{-\lambda t}}{1-e^{-\lambda T_0}} \qquad (2)$$

with the upper bound of time $T_0$, which is not uniform but exponential, quite irksome if we utilize it to yield random number directly. Owing to finite precision of the device in experiment, we have time least significant bit (LSB) bin size $t_0$ to digitize arrival time, with the maximum $N_0=T_0/t_0$. The probability to gain the value $k$ is given by (n is the discrete random variable):

$$P(n=k) = \int_{kt_0-t_0}^{kt_0} f(t)dt = \frac{e^{-\lambda t_0(k-1)} - e^{-\lambda kt_0}}{1-e^{-\lambda T_0}} \approx \frac{\lambda t_0 e^{-\lambda kt_0}}{1-e^{-\lambda T_0}} \qquad (3)$$

The last approximation in Eq. (3) is valid if $\lambda t_0 \ll 1$ which can be met as long as the mean number $\mu = \lambda T_0$ of detection events within $T_0$ is small enough and $N_0$ is big enough.

It is known from Eq. (3) that the probability decays exponentially as $k$ increases. Some people [8] try to shape $\lambda$ into $\lambda(t)$ by means of modulating the illumination intensity to change with time. If we let $\lambda(t) = 1/(T_0-t)$ by shaping the waveform, the distribution can be uniform in theory. However, the aforementioned $\lambda(t)$ cannot be well achieved physically. In order to register all detections, dead time $t_d$ should be as small as possible, which cannot be well achieved physically either. Even if it can be realized relatively well, a more serious problem arises that severe correlation has been brought in, for the sequence of $k$ within the same sampling interval is ordered from small to big. So in ref.[8], post-processing with SHA-256 is used for whitening the raw data.

If we only retain such periods as only one detection event appears, and then output $k$ as random numbers where $k$ is the time bin of period which has only one detection. Using conditional probability formula, we can prove simply that the distribution is uniform considering every time bin obeys Poissonian distribution with the mean $\lambda t_0$:

$$P_k = \frac{e^{-\lambda t_0}\lambda t_0(e^{-\lambda t_0})^{n-1}}{e^{-\lambda T_0}\lambda T_0} = \frac{t_0}{T_0} = \frac{1}{N_0} \qquad (4)$$

So every probability of every time bin is uniform. Likewise small dead time $t_d$ is required as well for experimental setup to register the number of detections within every sampling interval, thus we choose a PMT as our detector for its short dead time.

The width of electrical pulses from PMT output gives rise to the dead time $t_d$ of PMT. Poissonian distribution needs to be modified taking the effect of $t_d$ into consideration. The probability [16-17] of the incident that $m$ pulses turn up in a period of time $T_0$ is given by

$$P_{T_0}(X=m) = \frac{\mu_r^m}{m!}\sum_{i=0}^{K-m}\frac{(-\mu_r)^i}{i!}(1-(i+k-1)\frac{t_d}{T_0})^{m+i} \qquad (5)$$

with

$$\mu_r = \mu \cdot e^{-\mu\frac{t_d}{T_0}}, \quad K = \left\lceil\frac{T_0}{t_d}\right\rceil$$

The probability of only one pulse is given by

$$P_{T_0}(X=1) = \mu_r \sum_{i=0}^{K-1}\frac{(-\mu_r)^i}{i!}(1-\frac{it_d}{T_0})^{1+i} \qquad (6)$$

From another aspect two pulses will merge into one if the time between these two pulses is less than $t_d$. Therefore, the merged pulse width is longer than $t_d$. Nevertheless, we can merely register time points of rising edge of pulses limited by the performance of the device in the experiment, in which way we remain the periods within which only one pulse exists but not only one detection. Eventually the cases survive where the pulses of more than one detection fuse into one in that all intervals between two detections are shorter than $t_d$, which induces the deviation from theoretical analysis.

In our experiment, taking the dead time of our setup into more consideration, a detailed distribution is discussed below.

Thinking about the mechanism of the dead time $t_d$, we can easily have the thought that $t_d$ right before the period of interest needs to be analyzed together with the period $T_0$. If in the spell $t_d$ there is a detection arriving, the pulse induced by the detection will extend into the beginning short spell of the period in which another detection arrives to merge into one pulse of which the rising edge is in the spell $t_d$. And if no other detections arrive within the period, there is no pulse within the period. However, as a matter of fact the detection arriving in the beginning short spell of the period does exist, and cannot be registered due to $t_d$. Based on the discussion above, presumably, the possibilities of the smallest '$k$'s are a little smaller than the ones of the medium '$k$'s. We take two steps to analyze quantitatively the effect of $t_d$: first consider the case of no detections in the spell $t_d$; then consider the case that detections exist.

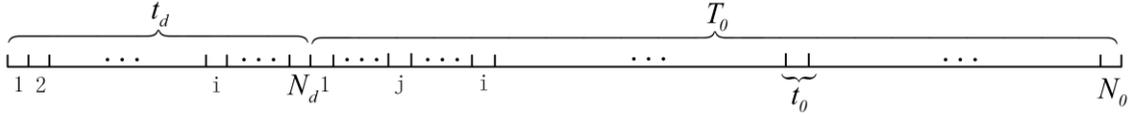

Fig.2 Schematic drawing of time bin of observed period and $t_d$.

As shown in Fig. 2, let $N_d = t_d / t_0$, period $T_0 = N_0 t_0$. If in the spell $t_d$ there is a detection arriving, the pulse induced by the detection will extend into the beginning short spell of the period in which another detection arrives to merge into one pulse of which the rising edge is in the spell $t_d$. And if no other detections arrive within the period, there is no pulse within the period. However, as a matter of fact the detection arriving in the beginning short spell of the period does exist, and cannot be registered due to $t_d$. Based on the discussion above, presumably, the possibilities of the smallest '$k$'s are a little smaller than the ones of the medium '$k$'s. We take two steps to analyze quantitatively the effect of $t_d$: first consider the case of no detections in the spell $t_d$; then consider the case that detections exist.

Although we retain the period of only one pulse, odds are that one pulse comprises several detections of which arbitrary pairs of two contiguous detections have intervals no more than $t_d$. We just calculate to the order of the combination of three detections for its single photon level light, and from the results below we can know the effect of three detections is rather minute. Define two more parameters, the probability of no detections in a time bin $p_0 = exp(-\lambda t_0)$ and the probability of the opposite case $p_e = 1 - p_0$. The probability of only one pulse is the sum of three terms, the probability of only one detection, that of two detections and that of three detections. It is easy to compute the first term that is independent from the serial number of time bin: $p_1(n,k) = p_0^{n-1} p_e$. The second term:

$p_2(n,k) = p_0^{n-2} p_e^2 \cdot h(n-k)$ where $h(n-k)$ denotes the number of the events of equal probability that the merged pulse occurs in the $k$th time bin. That is to say, the first detection is in the $k$th time bin and the second detection is less than $N_d$ later than the first, thus

$$h(x) = \begin{cases} N_d, & x > N_d \\ x, & 0 \leq x \leq N_d \\ 0, & x < 0 \end{cases} \quad (7)$$

Likewise, the third term: $p_3(n,k) = p_0^{n-3} p_e^3 \cdot h'(n-k)$ where $h'(k)$ is a bit more complex, but the nature is the same:

$$h'(x) = \begin{cases} N_d^2, & x \geq 2N_d \\ N_d(x-N_d) + (x-1)*(2N_d - x)/2, & \\ & N_d + 1 \leq x < 2N_d \\ x(x-1)/2, & 2 \leq x < N_d + 1 \\ 0, & x < 2 \end{cases} \quad (8)$$

From the above, the probability of only one pulse is given by $p(n,k) = p_1(n,k) + p_2(n,k) + p_3(n,k)$. The period $T_0$ in the experiment has $N_0 t_0$. Then we take the second step: assume that the last detection within the spell $t_d$ emerges in the $i$th time bin, the beginning $i$ time bins of the period is shadowed by $t_d$. If no detections are in the beginning $i$ time bins, then the probability of the pulse in the $k$th time bin of the period is $p(N_0 - i, k - i)$. And if detections are, assume that the last detection within the period emerges in the $j$th time bin (the serial number $j$ is reckoned from the beginning of the period), the next $N_d$ time bins are the $t_d$'s regime. Then we can continue to discuss based on whether there are detections during the spell, which can last to the end of the period. However, under a fairly faint illumination, the second order is precise enough. Therefore, after we add up the all possible $i$ and $j$, the probability that within the period only one detection appears and it is in the $k$th time bin is given by

$$P(N_0, k) = p_0^{N_d} \cdot p(N_0, k) + p_0^{N_d} p_e \cdot \sum_{i=1}^{N_d} p(N_0 - i, k - i) + p_0^{N_d} p_e^2 \cdot \sum_{i=1}^{N_d} p_0^{N_d - i} \sum_{j=1}^{i} p(N_0 - j - N_d, k - j - N_d) \quad (9)$$

Notably, if the second variable of $p(n,k)$ is less than or equal to 0, $p(n,k)$ should be 0. From the expression of $P(N_0,k)$, the approximation is equivalent to remaining the terms of equal to and less than three times of $p_e$. After the renormalization owing to just remaining the periods of only one pulse, the distribution is given by

$$P_{N_0}(n=k) = P_0(N_0,k) / \sum_{k=1}^{N_0} P_0(N_0,k) \qquad (10)$$

Eq. (10) is pretty hard to be simplified, but by means of some mathematical softwares like Mathematica, we can reap the numerical outcomes when $\lambda t_0$ is given as shown red solid line in Fig.3.

## 3. EXPERIMENTS AND RESULTS

The setup of experiment is shown in Fig.1, The precision of TDC is 0.162ns and the width of PMT pulses is about 5ns and the observed period $T_0 = N_0 t_0 = 320*0.162ns = 51.84ns$. The experimental data and theoretical values computed are drawn in the same plot as shown in Fig.3.

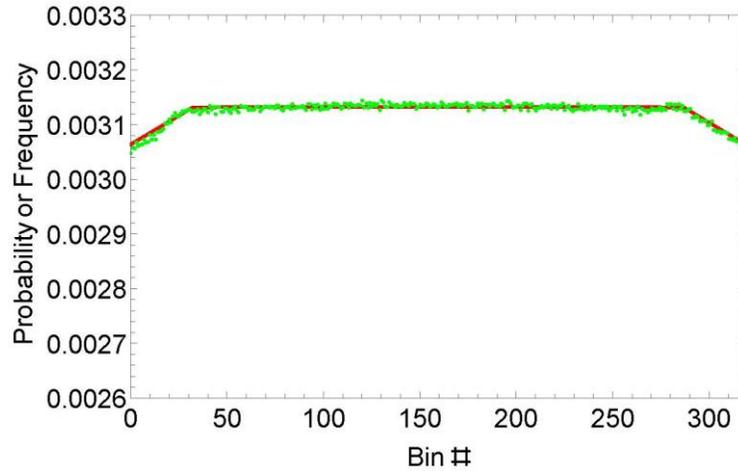

Fig.3 Diagram of comparison of experimental result and theoretically analysis. Red solid line is drawn theoretically based on Eq. (10) and green dots are experimental data points. The experimental data dots show a good agreement with the theoretical predictions

Let $\lambda t_0 = 0.00068$ to fit the scatter plot, thus the mean number $\mu$ of photons within a period is equal to $\lambda T_0 = 0.2176$. Meanwhile, the dead time $t_d \approx N_d t_0 = 32*0.162ns = 5.184ns$ to make $t_d / T_0 \approx 0.1$.

The min-entropy can be calculated according to Eq. (10):

$$H_\infty = -\frac{Log_2 P_{max}}{Log_2 N} = -\frac{Log_2 0.003132}{Log_2 320} = 0.99961$$

which is not great enough, thus we cut off $N_d$ on the both sides, renormalize it, and the min-entropy is given by

$$H_\infty = -\frac{Log_2 P_{max}}{Log_2 256} = -\frac{Log_2 0.00390633}{Log_2 256} = 0.999996 \qquad (11)$$

which is as good as the min-entropy of Ref. [9] and can be passed any tests. The remaining proportion after cutting can be given by as well

$$\sum_{k=N_d+1}^{N_0-N_d} P_{N_0}(n=k) \approx \frac{(T_0 - 2t_d)}{T_0} \qquad (12)$$

Using Eq. (6), Eq. (10) and Eq. (12), every period needs to cut off $t_d$ on the both ends and the bitrate is approximately given by

$$v = Log_2(\frac{T_0 - 2t_d}{t_0}) \cdot P_{T_0}(X=1) \cdot \frac{1}{T_0} \cdot \sum_{k=N_d+1}^{N_0-N_d} P_{N_0}(n=k) \approx Log_2(\frac{T_0 - 2t_d}{t_0}) \cdot P_{T_0}(X=1) \cdot \frac{1}{T_0} \cdot \frac{T_0 - 2t_d}{T_0} \qquad (13)$$

with $N_d = t_d / t_0$. The first term is the bit number of $k$; the second one is the probability of only one pulse; the third one is frequency; the forth one is the proportion of the data left after truncation that is roughly represented by the ratio of intercepted time to the whole period as a result of the quasi-uniform distribution. From Eq. (5) $P(X=1) = 0.1786$, and using the parameters above associated with Eq. (13) the bitrate $v \approx 8*0.1786*(1/51.84ns)*0.803 \approx 22.13Mbps$ which tally well with the one in the experiment. If we improve the illumination intensity continuously on condition, random numbers are generated at a rate up to 45Mbps that passed the prevalent tests.

The real min-entropy cannot attain such a level as a result of a variety of imperfections of devices, especially the counting loss of TDC. Taking the frequency of the data as the probability to calculate min-entropy, we can obtain $H_\infty = 0.99900$ from the maximum of the frequency 0.003143, regarded as a reference of the min-entropy of the real distribution. Cutting off $N_d$ on the both ends, we can increase the min-entropy as shown in Fig.3 which exhibits the practical measured data with green dots line. The experimental data dots show a good agreement with the theoretical predictions and the effect of eliminating the periods of more than one detections is clearly verified by the comparison between two curves. After cutting off $N_d$, the remaining proportion $0.803 \approx 256/320$ proves the reasonability of the approximation of Eq. (13). According to Eq. (10), the min-entropy $H_\infty = 0.999996$ is perfectly enough to be applied to direct practice. Likewise, the real min-entropy $H_\infty = 0.99946$ can be derived from the biggest frequency 0.003918.

From Eq. (13) it is apparently shown that $T_0$, $P_{T_0}(X=1)$, $t_d$ and $t_0$ all have a direct effect on the bitrate. $T_0$ and $P_{T_0}(X=1)$ can be easily controlled by adjusting the logic of FPGA which used in our experiment and the illumination intensity respectively. Based on the premise of uniformity distribution, the two parameters should be set to lift the bitrate as fast as we can. $P_{T_0}(X=1)$ can be tailored more than 0.4. $t_d$ and $t_0$ are determined by the performance of the experimental devices and cannot be changed.

From the discussion above, the core factor is the dead time $t_d$. The ratio of $t_d$ to $T_0$ should be small enough, generally set at 0.1 or less in the experiment, otherwise refined uniformity will be lost, which restrains the lower bound of $T_0$. Halving $t_d$ means we can halve $T_0$ to double the bitrate, while halving $t_d$ cannot improve the bitrate significantly resulting from the logarithm calculation of the term. Generally we set eight bits for the sake of being processed easily, and meanwhile to ensure the suitable ratio of $t_d$ to $T_0$.

There may be a sense that we have discarded a large part of the data in that the periods of more than one detection are rejected. From Eq. (12), ratio of the remaining data is great than 0.8. Thus our scheme still keeps a high utilization level for the raw sequence.

Another most notable merit of our scheme is that the perfectly uniform distribution after cutting off $N_d$ on the both sides depends little on the illumination intensity. Thus the quality of the random numbers generated in the experiment was extremely steady and no apparent deviation happened in the whole experimental process. Unlike the scheme of Ref. [11], an observable bias can be caused by a minute drift of the illumination intensity, so real-time controlled needs doing. Ref. [18] designed a mathematical method called Beyasian estimation for the purpose of a better estimation and manipulation. Our scheme avoids these troubles simply and thoroughly.

Here the effect of dark counts and afterpulses are discussed. In our setup in Fig.1, the dark count of PMT is very small which is about several hundred per second we have measured. So we can ignore its influence as the photon pulse's frequency is over millions of counts per second. When a PMT is operated in a pulse detection mode, spurious pulses with small amplitudes may be observed which is called afterpulses. We know the amplitude of afterpulses is small and most of them are less than 10% of true pulses [19]. In our setup, we set a threshold of discriminator more than amplitude of noises and afterpulses, and eliminate the influence of the afterpulses.

## 4. RANDOMNESS TESTS

Two most famous tests are used which are the "Statistical Test Suite" (STS) [20] and the "DieHard" (Die) test suite [21], to test our random numbers. The results are shown in Fig.4. Every test is performed many times to give out multiple *p-values* to which $\chi^2$ *Test* is applied finally in STS while in Die *Kolmogorov-Smirnov Test* [22] is applied. We adopted the same operating parameters and quantity as before for Die, while for STS the total of *1Mbit * 1000* was tested with built-in parameters. It is notable that several tests are operated more than once with different parameters, especially the test of *NonOverlappingTemplate* 148 times. For the sake of better exhibition we apply *Kolmogorov-Smirnov Test* to multiple *p-values* and display mean values for multiple proportion values.

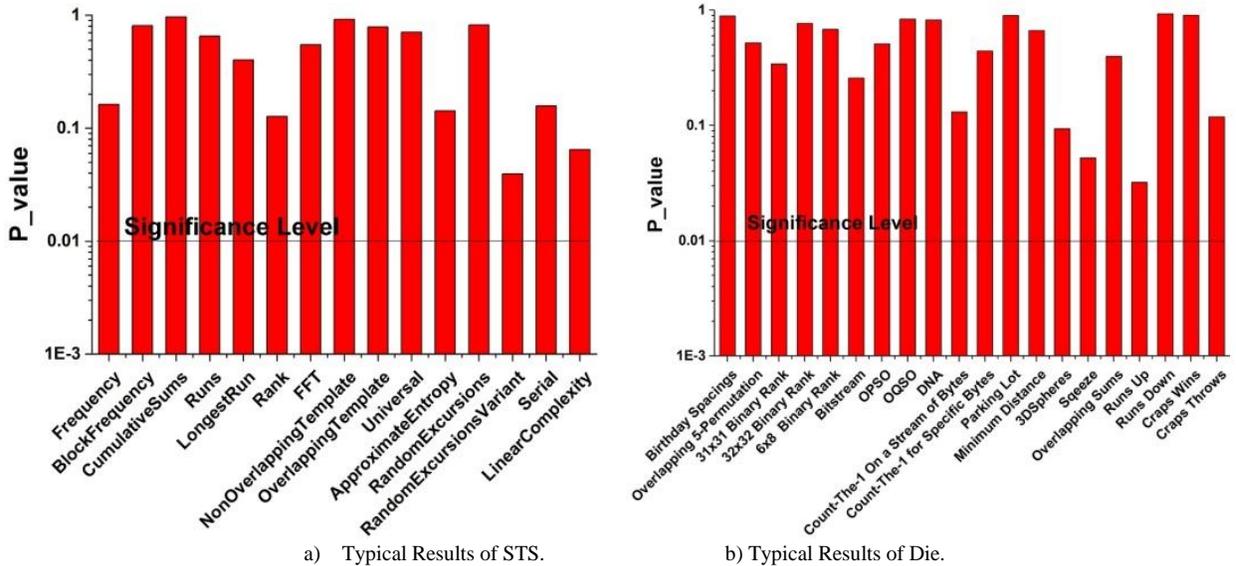

a) Typical Results of STS.    b) Typical Results of Die.

Fig.4 Typical results of STS test and DieHard test. Without post-processing, the p-values are routinely above the significance level confirming the quality and the reliability of the QRNG

About 100GB data in total generated in all experimental processes are analyzed through the tests. No evident defects have not been found, which sufficiently show the superiority of our scheme.

## 5. Conclusions

We transform the obnoxious exponential distribution into desirable uniform distribution through our simple scheme. We accomplished the experimental verification for our theoretical analysis using our simple devices and achieved the steady output of high-quality and ready-for-use random numbers. Furthermore, our scheme is nonparametric more stable than any schemes with parameters involved. The bitrate can be improved largely, influenced by the core parameter $t_d$. If $t_d$ can arrive at 1ns (The rise time of today's PMT is about 0.5ns that is, two detections of which the interval is about 0.5ns can be resolved as long as the circuits following PMT have a high enough speed), and TDC has a higher precision like 10ps, the bitrate of our QRNG can go up to 320Mbps or so.

## Acknowledgements

We acknowledge the financial support from the Fundamental Research Funds for the Central Universities, National Natural Science Funds of China under Grant No: 11178020, 11275197 and the CAS Special Grant for Postgraduate Research, Innovation and Practice.

## References

[1] J. H. Davenport, Papers from the international symposium on symbolic and algebraic computation Pages, ISSAC 123-129 (1992).
[2] A. J. Menezes, P. C. van Oorschot and S. A. Vanstone, Handbook of Applied Cryptography (CRC Press, London, 1997).
[3] N. A. Frigerio and N. Clark, A random number set for Monte Carlo computations, Trans. Amer. Nucl. Soc., 22 (1975), pp. 283-284.
[4] A. Stefanov, N. Gisin, L. Guinnard, H. Zbinden, "Optical quantum random number generator," J. Mod. Opt. 47, 595-598 (2000).
[5] J. G. Rarity, P. C. M. Owens, and P. R. Tapster, "Quantum random-number generation and key sharing," J. Mod. Opt. 41, 2435-2444 (1994).
[6] T. Jennewin, U. Achleitner, G.Weihs, H.Weinfurtur, A. Zeilinger, "A fast and compact quantum quantum random number generator," Rev. Sci. Instr. 71, 1675-1680 (2000).
[7] M. A. Wayne, E. R. Jeffey, G. M. Akselrod, and P. G. Kwait, "Photon arrival time quantum random number generation," J. Mod. Opt. 56, 516-522 (2009).
[8] M. A. Wayne, and P. G. Kwait, "Low-bias high-speed quantum random number generator via shaped optical pulses," Opt. Express 18, 9351-9357 (2010).
[9] M. Stipčević, and B. Medved Rogina, "Quantum random number generator based on photonic emission in semi semiconductors," Rev. Sci. Instrum. 78, 045104 (2007).
[10] M. Wahl, M. Leifgen, M. Berlin, T. Röhlicke, H. Rahn, and O. Benson, "An ultrafast quantum random number generator with provably bounded output bias based on photon arrival time measurements," Appl. Phys. Lett. 98, 171105 (2011).
[11] M. Fürst, H.Weier, S. Nauerth, D. G. Marangon, C. Kurt-siefer, and H.Weinfurter, "High speed optical quantum random number generation," Opt. Express 18, 13029-13037 (2010).
[12] Z. L. Yuan, M. Lucamarini, J. F. Dynes, B. Fröhlich, A. Plews, and A. J. Shields, "Robust random number generation using steady-state emission of gain-switched laser diodes," arXiv.org 1407.0933 (2014).
[13] B. Sanguinetti, A. Martin, H. Zbinden, and N. Gisin, "Quantum random number generation on a mobile phone," arXiv.org 1405.0435 (2014).
[14] J. F. Dynes, Z. L. Yuan, A. W. Sharpe, and A. J. Shields, et al. "A high speed, post-processing free, quantum random number generator." applied physics letters 93.3 (2008): 031109
[15] Mario Stipčević and John Bowers, Post-Processing Free Spatio-Temporal Optical Random Number Generator Resilient to Hardware Failure and Signal Injection Attacks, arXiv:1410.0724
[16] K.Omote, "Dead time effects in photon-counting distributions," Nucl. Instr. Methods 293, 582-588 (1990).
[17] J. W. Mueller, "Some formulae for a dead-time-distorted poisson process," Nucl. Instr. Methods 117, 401-404 (1974).
[18] P. Lougovski, and R. Pooser, "An observed-data-consistent approach to the assignment of bit values in a quantum random number generator," arXiv.org 1404.5977 (2014).
[19] U Akgun, A S Ayan, G Aydin, F Duru, J Olson and Y Onel, "Afterpulse timing and rate investigation of three different Hamamatsu Photomultiplier Tubes", 2008 JINST 3 T01001 doi:10.1088/1748-0221/3/01/T01001.
[20] J. Soto, "Statistical testing of random number generators," Proc. 22nd National Information Systems Security Conference (1999).
[21] R. G. Brown, "Dieharder test suite," http://www.phy.duke.edu/ rgb/General/dieharder.php (2009).
[22] N. H. Kuiper, "Tests concerning random points on a circle," Proc. Kon. Ned. Aka Wet. A 63, 38-47 (1962).